\documentclass[12pt]{article}

\usepackage{graphicx}
\usepackage{url}
\begin{document}

\def\b{\bigskip}
\def\s{\smallskip} \def\c{\centerline}\def\r{\rangle}\def\i{\int}
\def\e{\lambda} \def\z{\theta}

\b
\c {\bf Decoherence and  Planck's 
Radiation Law }\b 

\c {Italo Vecchi} \c {Bahnhofstr. 33 - 8600 Duebendorf -
Switzerland} \c { email: vecchi@isthar.com }\b

In the present  note the foundations  of the theory of environment-induced
decoherence  are considered. It is pointed out that the common arguments
 on the diagonalisation 
of the density matrix are based on questionable hidden 
assumptions, conflicting with accepted physical results. 
An alternative interpretation of the phenomeana related 
to decoherence is proposed.  We
refer to (Joos 1999) and (Zurek 1999) for introductions to the subject of decoherence
theory and for background material. An historical perspective of
decoherence theory with some relevant critical remarks on the
diagonalisation process is provided in (Santos \& Escobar, 1999).\s 

We focus on Joos´ excellent survey of decoherence theory (Joos 1999), whose
clarity makes its relatively easy to spot the inconsistencies in the
argument.  Given a system $ S $ in a
superposition of eigenstates $ |n \rangle $ and its environment $ W $
in a state $ \Phi_o $, the pointer states are identified as those states
$|\Psi(t)\rangle$ in $ W $ resulting from the interaction between $ S $
and $ W $ $$ |n\rangle |\Phi_o\rangle \longrightarrow exp(-iH_{int})
|n\rangle |\Phi_o =: |n\rangle |\Phi_n(t)\rangle. $$ The states $
|\Phi_n(t)\rangle $ result from the entanglement of the environment $ W $ with $ S $ through the
interaction Hamiltonian $ H_{int} $ and are usually referred to as the
"pointer positions". In this setting the environment $ W $ includes any macroscopic measurement device, which is assumed to be strongly coupled to the rest of the universe.
An act of measurement on $ W $ induces a collapse
of its state vector into one of the pointer's vector, yielding
information about the state of the system $ S $. The states $
|\Phi_n(t)\rangle $ are described in (Joos 1999) as the states of the "rest of
the world". According to decoherence theory the density matrix relative to  $|\Psi(t)\rangle$ is rapidly 
reduced to a diagonal form, reflecting the system's  entanglement with the environment. The off-diagonal interference 
terms in the density matrix   
vanish, as superpositions become inaccessible  to local observers.\s

 The basic ambiguity underlieing this description of the
decoherence process stems from the fact that any vector basis can
be chosen as a pointer or "preferred" basis, so that the very concept of pointer basis is ambiguous.
Since the environment and any measurement device
can be described using an arbitrarily chosen basis $|\Psi(t)\rangle$,
the "preferred" pointer basis referred to by Joos can only be  relative to an
observer, as defined by a measurement operator. \s

It should be clear that the measurement device
or the environment do not chose a basis, as physical systems do not chose reference systems. The observer does. The
privileged pointer basis is actually  determined by the set of possible outcomes
of a measurement act performed by an observer. It is the intervention of
the observer on the measurement apparatus or on the environment  in the course of the
measurement process that determines the pointer basis. \s 

An example may clarify the underlying issue. The black body radiation is an instance of macroscopic phenomenon 
described by a well-understood quantum model.  Planck's radiation law 
$$ \rho(\omega, T) = 
  { 1 \over { \pi^2 c^3} }    
{ {\hbar \omega^3} \over {exp(\hbar \omega / k_B T) -1 }} $$ is obtained maximising entropy on
discrete energy spectra.  In the black body model the evolution of the radiation field is a continuous, reversible process governed by  Schroedinger equation that induces the smooth 
evolution of the system's state vector.
On the other hand entropy is maximised on
discrete energy spectra. The equilibrium distribution
$$ f = { 1 \over {  exp(\hbar \omega / k_B T) -1 } } $$ 
is obtained (Planck 1900 cf. Kuhn 1978, Mackey 1993) maximizing the 
Boltzmann-Gibbs entropy $ G(f) = -\int f log f $ on the discrete set 
$ \epsilon_n = n \hbar \omega \quad ; n = 0,1,2, ... . $, i.e. on the eigenvalues
of the  energy operator $ H $.
Planck's law is then obtained as a product of $ f $ and of the mode density 
$ { { \omega^2 } / { \pi^2 c^3} } $.  Entropy maximisation
may be applied to other
sets of observables too, but it will yield different results. Entropy maximisation 
on continuous spectra yields the Jeans-Raleigh law (Einstein 1906, cf. Kuhn 1978). 
Other observables yield other
distribution laws ( Mackey 1993). \s

Planck's radiation law depends on a "preferred basis", i.e on the set of eigenvectors of
the energy operator.
In other words the Planck distribution is obtained 
maximizing the entropy of a set
of energy measurements, i.e. maximizing the observer's lack of information  on
the measurement outcomes of the energy operator.
According to decoherence theory the emergence of a "preferred" basis, such as the energy basis in the case of Planck's radiation law,  is induced by interaction with the environment. 
However Planck's  radiation law  applies also to "closed" systems, so that the emergence of the "preferred basis" relative to the energy operator cannot be attributed to entanglement resulting from interaction with the environment. A concrete example is the cosmic background radiation, which complies with Planck`s radiation law in the absence of any external environment, the radiation field being decoupled from the electrically charged matter. \s

In general the density matrix $ D $ corresponding to $f$ is $$ 
 D = { e^{-H \over kT} \over  Trace( e^{-H \over kT}) } $$
so that its off-diagonal elements are null in the energy eigenbasis
(Von Neumann 1932, V.3). The fact that the  
the off-diagonal elements of $ D $ are null however does not depend  on any interaction
 with the environment, since the system may well be isolated.
Actually if the system is isolated its evolution is unitary 
so that its state $\Psi$ is a pure state yielding a density matrix
 $G=\Psi\otimes\Psi$ with non-null  diagonal elements, which however are generally unknown to to the observer.
The above distinction between $ G $ and $ S $ is essentially the same as that
between $ {\sl type1}$ and $ {\sl type2}$ systems in quantum information theory.
 ( see Pospiech 2000 for a survey). \s

If we try to interpret this process in terms of environment-induced decoherence we can spot where the key misunderstanding about the meaning of the density matrix arises.
The matrix $ D $ where the off-diagonal elements are null 
just does not represent the state of the system 
but only encodes the observer's knowledge of the measurement outcomes relative to the energy operator.
In the language of quantum measurement theory the matrix $ D $ refers to a  mixture. The fact that the off-diaginal elements of the matrix $ D $ are null is hence seen to depend on the observer, as defined by
a set of observables or, equivalently, by a measurement operator.
The increase the system's entropy just reflects the loss of information of the observer associated 
to a measurement operator. \s

It worth rememembering that in general the property that a  density matrix  $ S $ describing a mixture
is diagonal with $ Trace(S) = 1 $ encodes only trivial information
on the fact that the measurement will yield some result. Non-trivial diagonal information, i.e. 
non-trivial information on measurement outcomes, is encoded in the specific values of the 
diagonal elements.
In the case of the black body the  macroscopic information determinining
 the values of the diagonal elements
is provided by  conservation of energy, by the temperature and by the properties of the 
energy spectrum.\s 

The role of the observer in the decoherence argument is indeed
acknowledged in (Joos 1999), as is the fact that the superpositions in the
system are not destroyed but merely cease to be identifiable by local
observers. However in decoherence theory the pointer basis is implicitly treated as an
intrinsic property of the interaction between the system and its
environment or a measurement device. This tacit assumption is necessary
for the environment-induced decay of the off-diagonal interference terms of the system's
density matrix, $$ \rho_S = \sum_{n,m} c^*_m c_n |m\rangle \langle
n|\longrightarrow \rho_S = \sum_{n,m} c^*_m c_n \langle \Phi_m| \Phi_n
\rangle |m\rangle \langle n| $$ which is then interpreted as the
vanishing of superpositions. The assumption however leads to
inconsistencies, as shown by the following analysis.\s

Treating the pointer basis as an intrinsic property of the environment would not
matter if the decoherence argument was independent of the chosen pointer
basis. However this is not the case. According to the argument in (Joos 1999)
and (Zurek 1993) , the decoherence process induces the decay of the off-diagonal
elements of the systems density matrix, $$ \rho_S ´\longrightarrow
\sum_n |c_n|^2 |n\rangle \langle n| $$ which is interpreted as the
emergence of a set of stable macroscopic states. The density matrix
however is defined in terms of the pointer basis. Different pointer
basis lead to different density matrices for the same state vectors. It
is immediate to see that the decoherence process, i.e. the decay of the
off diagonal terms in the density matrix, does not commute with a change
of basis. Indeed given a density matrix $ A $ , let $ C $ be a change of
basis and , $ C^{-1}$ its inverse and D the operator that equates to
null the off-diagonal elements. Then $$ DA \neq (C^{-1} D C) A $$ so
that the result of the decoherence process depends on the pointer basis,
which is selected by the observer and is independent of the underlying
physical process. The states associated with a
diagonal density matrix in one basis describe superpositions in the
other basis. Indeed any two non-commuting operators induce pointer
basis for which the above inequality holds, so that the physical process inducing the diagonalisation appears to depend 
on the chosen basis.  This is absurd, unless one accepts that the diagonal matrix
describes an observer-dependent mixture, for which the above argument does not hold.
The root of the mistake is the attempt to "objectify" the
observer's loss of information, attributing it to a physical process unrelated to the observer.\s

 The above indicates that
the result of the entropy maximisation process depends on the observer
and that it applies to the measurement outcomes relative to the observer's
measurement operator. If our interpretation is correct 
 there must be a flaw in the  argument tieing the decay of 
the off-diagonal elements of the density matrix to the interaction with the environment.\s

The flaw is not hard to find. If one examines the
argument leading to the diagonalisation of the system's density matrix,
one discovers that it is based on unphysical no-recoil assumptions on
the scattering process (Joos 1999), i.e. on ignoring  back-action on the environment either directly
or through appropriately chosen cut-offs (cf. Unruh \& Zurek 1989) or through
selective application of fine/coarse graining to different variables (Brun 1993, cf.
Feynman \& Vernon 1963). It may be noted that the fine/coarse graining approach reveals
the role of the observer, which was later fudged by uncritical use of the original results. 
Under the  no-recoil assumption every scattering
event multiplies the off-diagonal elements of the local density matrix by a factor
$ 1-\epsilon $ (Joos 1999, 3.1.2).  This hammers the non-diagonal elements into converging to zero, while
preventing the environment from eroding the diagonal elements. 
The no-recoil assumption forces the density
matrix into a very singular form, where the off-diagonal terms converge
rapidly to zero, while the diagonal terms remains intact. Applying the
no-recoil assumption to a different basis however leads to a diagonal
matrix describing a different physical state and which is not diagonal
under a change of basis, as shown above. \s

On the other hand, as shown by the Planck's radiation law, 
a diagonal matrix referring to a mixture
 is naturally associated to the system, not on the basis of any physical 
interaction with the environment,
but simply on the basis of entropy maximisation of the mixture relative to a measurement operator. Such entropy maximisation 
yields the saystem's macroscopic properties relative to the observer associated to the operator.\s 

The  decoherence process reflects then the
observer's loss of information, not only on superpositions, but on the microscopic
state of the system. The special status of superpositions is indeed
spurious, since it depends on the measurement operator being considered,
i.e. on the observer. The singling out of superpositions, i.e. of off-diagonal
elements of the local density matrix,  for special
destructive treatment appears as an artefact, based on
unphysical assumptions and on confusion  between $ {\sl type1}$ with $ {\sl type2}$ systems, 
i.e. on attributing  pure states' properties to mixtures.\s

We wrap up our considerations with a simple "Schroedinger's cat" example,  illustrating
the constraints of global unitarity on local observers.
Consider the situation $$ System = Cat ,\qquad   Environment = Rest\ of\ the\ World. $$ 
and the basis $ A= (|alive> ,  |dead> )$.
The system's initial state is $ {1 / \sqrt 2} ( |alive> +  |dead> )$.
We may  consider the system in the basis
$$ B= (B_1, B_2)=({1 / \sqrt 2} (|alive> + |dead>), 
{ 1 / \sqrt 2} (|alive> - |dead> )). $$
A change of basis does not affect the state of the system,
as long as no basis-dependent measurement takes place. 
As long as the Cat is not observed, 
the universe's state-vector, whose evolution is unitary,  
is $$ B_1^{universe} = {1 / \sqrt 2} (|alive^{universe}(t)> + |dead^{universe}(t)>). $$
The phase-related information about the Cat-superpositions is encoded in $ B_1^{universe}(t) $ 
and it may not be accessible to a basis-$A$-observer in the state-of-the-Cat subsystem,
which can be represented either as a mixture by a diagonal matrix ($type2$ system) reflecting the
observer's ignorance in a specific basis, or as non-diagonal density matrix with unknown 
non-diagonal elements ($type1$ system). For basis-$A$-observers
the Cat will either die or live once the Cat-subsystem is projected 
onto that basis by an act of observation/measurement.
For an hypothetical observer in basis $B$ however
there are no superpositions. The state-of-the-Cat density-matrix in basis $B$
is just  $$ \pmatrix { 1 & 0 \cr 0 & 0 } $$ and the outcome of a basis-$B$-measurement
is certain. Such tilted-basis measurements are actually at the core of the Elitzur-Vaidman 
scheme (Elitzur \&  Vaidman 1993), where information is extracted from a system without inducing collapse in the "usual" basis. 
 
\b \vfill\eject
\baselineskip=12 pt

 \c{ \bf References} \b

 Brun T.A. 1993 Quasiclassical Equations of Motion for Nonlinear Brownian Systems
{\sl Phys. Rev. D} 47 pp 3383-3393.\s
 
 Einstein A. 1906 Zur Theorie der Lichterzeugung und Lichtabsorption
{\sl Annalen der Physik} 20 pp 199-206.\s

 Elitzur A. and  VaidmanL. 1993 Quantum Mechanical Interaction-free Measurements {\sl  Found. Phys.} 23(7) pp 987-997. \s

 Feynman R.P. \& Vernon F.L. 1963 The Theory of a General Quantum System
Interactibg with a Linear Dissipative System
{\sl Ann. Phys. (NY) } 24 pp 118-173.\s  

 Mackey M.C. 1993 Time's Arrow: The Origins of Thermodynamic Behaviour. 
 Spinger.\s

 Joos A. 1999 Decoherence Through Interaction
with the Environment. In {\sl Decoherence and the Appearance of a Classical
World in Quantun Theory} ( ed. D.Giulini et al.) pp 35-136, Spinger.\s

 Kuhn T.S. 1978 Black-Body Theory and the Quantum Discontinuity, 1894-1912. 
 The University of Chicago Press.\s
 	
 Paz J.P. \& Zurek W.H. 1999 Quantum
limits of decoherence Environment induced superselection of energy
eigenstates. {\sl Phys. Rev. Lett.} 89, pp 5181-5185.\s 

 Planck M. 1900 Zur Theorie des Gesetzes der Energie Verteilung im Normalspectrum.
{\sl Verhandlungen der Deutschen Physikalischen Gesellschaft} 2 pp 237-245.\s
 
 Pospiech 2000 Information - the Fundamental Notion of Quantum Theory {\sl \vskip 0.0pt \noindent
at http://xxx.lanl.gov/abs/quant-ph/0002009 } \s

 Santos L. \& Escobar C.0.  1999 Convergences in the Measurement Problem in Quantum Mechanics 
{\sl at http://xxx.lanl.gov/abs/quant-ph/9912005} \s

 Unruh W.G. \&  Zurek W.H. 1989 Reduction of a wave packet in quantum Brownian motion  
{\sl Phys. Rev.} D40 pp 1071-1094.

 Von Neumann J. 1932 Mathematische Grundlagen der Quantenmechanik
Springer.\s

 Zurek W.H. 1993 Preferred Observable of Predictability, Classicality and the
Environment Induced Decoherence. {\sl Progr. Theor. Phys.} 89 pp 281-312. 

\end{document}